\documentclass[twocolumn,secnumarabic,amssymb,amsmath,superscriptaddress,aps,prl,longbibliography]{revtex4-1}

\usepackage{filecontents}
\usepackage{graphicx}
\usepackage{amsfonts}
\usepackage{amssymb}
\usepackage{bm}
\usepackage{color}
\usepackage{xspace}
\usepackage{xr}
\newcommand*\ii{\mathrm{i}}
\newcommand*\rmv{\mathrm{v}}
\newcommand*\com{c.m.\xspace}

\newcommand*\bohr{\xspace$a_0$\xspace}
\newcommand*\au{\xspace}
\newcommand*\figurewidth{0.96\columnwidth}
\allowdisplaybreaks

\begin{document}

\title{Dissecting strong-field excitation dynamics with atomic-momentum spectroscopy}

\author{A.W.\ Bray}
\email{alexander.bray@anu.edu.au}
\affiliation{Australian National University, Canberra}
\affiliation{Max-Born-Institute, Berlin, Germany} 
\author{U.\ Eichmann}
\email{eichmann@mbi-berlin.de}
\author{S.\ Patchkovskii}
\email{serguei.patchkovskii@mbi-berlin.de}
\affiliation{Max-Born-Institute, Berlin, Germany} 

\date{\today}

\begin{abstract}
Observation of internal quantum dynamics relies on correlations between the
system being observed and the measurement apparatus. We propose using the
center-of-mass (\com) degrees of freedom of atoms and molecules as a
``built-in'' monitoring device for observing their internal dynamics in
non-perturbative laser fields. We illustrate the idea on the simplest model
system - the hydrogen atom in an intense, tightly-focused infrared laser beam.
To this end, we develop a numerically-tractable, quantum-mechanical treatment
of correlations between internal and \com dynamics.  We show that the
transverse momentum records the time excited states experience the field,
allowing femtosecond reconstruction of the strong-field excitation process. The
ground state becomes weak-field seeking, an unambiguous and long sought-for
signature of the Kramers-Henneberger regime.
\end{abstract}

\maketitle

The process of measurement in quantum mechanics relies on establishing a
correlation between an internal quantum degree of freedom and a classical
degree of freedom of a measurement apparatus. Finding a suitable classical
outcome for a quantum system of interest is particularly important for
achieving optimal temporal and spatial resolution. One classical degree of
freedom available to every gas-phase system is the translational motion of its
center of mass (\com), effectively attaching an individual measurement
apparatus to each atom or molecule. The closely-related prescription of using
the \com motion as a control device has been very successful in
M\"ossbauer\cite{Greenwood71a} and other Doppler spectroscopies\cite{Klaft94a}.

The coupling between the internal quantum dynamics and the \com motion has not
received much attention in strong-field atomic, molecular, and optical (AMO)
science.  In intense visible and infra-red fields, this coupling is a subtle
effect, intimately connected to the breakdown of the dipole approximation. The
fundamental importance  of non-dipole effects have been recognized early
on\cite{Reiss90a,ChL:08,Rei:08}, but only recently, enabled by refined
theoretical and experimental approaches, processes beyond the
dipole-approximation  are coming into focus.  These include radiation
pressure\cite{LMM:14}, momentum distribution between fragments upon
ionization\cite{SAZ:11,CBC:14,HEB:19}, chiral effects in HHG\cite{Cireasa15a},
and atomic acceleration\cite{ENR:09}.  These effects have been investigated for
very intense (relativistic and near-relativistic) infra-red (IR)
fields\cite{DDE:01,KYB:13,KHW:17,PDC:05}, as well as for shorter-wavelength
fields which are becoming available in the strong-field regime\cite{FoS:16}.

Because the \com coupling effects in strong-field physics are small, numerical
treatment of their contribution is challenging. The standard technique appears
to be the treatment on full-product grids\cite{Chelkowski15a}, which would
require a 6D numerical simulation even for the simplest realistic target -- the
hydrogen atom.  

In this Letter we show that adding an artificial trapping potential, chosen not
to disturb the \com motion, allows the effective dimensionality of the problem
to be reduced to 3D.  This enables detailed computational investigation of \com
dynamics of strong-field processes.  By using the \com motion as the
``built-in'' measurement apparatus, we obtain information on the dynamics of
the excited-state formation in intense IR fields. Using this technique, we
provide the first unambiguous, experimentally-realizable method for confirming
the atomic ground state transiently entering the Kramers-Henneberger (KH)
regime in such fields.

In the KH (or acceleration) frame of reference, the laser field dominates the
electronic motion. For a laser field with the peak electric field amplitude
$F_0$ and carrier frequency $\omega$, linearly polarized along the direction
$\hat{n}$, the lowest-order Fourier component of the interaction potential in the
KH frame takes the form\cite{Henneberger68a}:
\begin{align}
   U_{\rm KH}\!\left(\vec{r}\right) &= \frac{1}{2\pi} \int_0^{2\pi} 
     U\!\left(\vec{r}+\vec{l}_0 \sin\!\left(\tau\right)\right) d\tau, & \label{eqn:vkh}
\end{align}
where $U$ is the interaction potential in the laboratory frame and the electron
oscillation amplitude $\vec{l}_0=\hat{n} F_0 \omega^{-2}$.

If higher-order corrections to Eq. \eqref{eqn:vkh} can be neglected for a given
state, the system is said to be in the Kramers-Henneberger regime.  A
remarkable property of the KH states in low-frequency fields is that the effective
polarizability rapidly approaches $-\omega^{-2}$\cite{WWK:17} with increasing
$\vec{l}_0$ magnitude. As the result, a system in a KH state experiences the
same ponderomotive potential as a free electron.

Kramers-Henneberger states have been postulated to explain photoelectron
spectra in strong fields\cite{Morales11a}, ionization-free filamentation in
gases\cite{Richter13a}, and ponderomotive acceleration of neutral excited
states\cite{ENR:09,EZE:14,ZiE:16,WWK:17,STM:17,ZMK:18}. Rydberg states readily
satisfy the KH criteria in intense IR fields, and are commonly accepted to be
in the KH regime in such fields. Because the KH states exist only
transiently in the presence of the intense field, their unambiguous
detection remains elusive\cite{WWK:17}. The mechanism of their formation in
low-frequency fields, and for the ground state even their existence, remain
controversial\cite{Popov99a,Gavrila02a,Popov03a,Simbotin04a,Gavrila08a},
despite extensive
investigation\cite{deBoer92a,JSB:93,NGS:08,Wolter14a,PMM:17,OHL:18,CGL:12,Li14a,ZPI:17}.


In the simplest case of a 1-electron, neutral atom, the laboratory-frame
Hamiltonian is given by (unless noted otherwise, atomic units ($\hbar\!=\!m\!=\!|e|\!=\!1$)
are used throughout):
\begin{align}
  \hat{H} & = \frac{1}{2m_1} \left(\hat{p}_1 + \vec{A}\!\left(\vec{r}_1,t\right)\right)^2
            + \frac{1}{2m_2} \left(\hat{p}_2 - \vec{A}\!\left(\vec{r}_2,t\right)\right)^2 \nonumber \\
          & + v\!\left(\vec{\chi}\right) + u\!\left(\vec{R}\right) & \label{eqn:h-lab}
\end{align}
where $\hat{p}_{1,2}$ are the momentum operators of particles $1$ (electron,
charge $q_1\!\!=\!\!-1$) and $2$ (nucleus, $q_2\!\!=\!\!+1$),
$\vec{A}\!\left(\vec{r},t\right)$ is the transverse
($\hat{\nabla}\cdot\vec{A}=0$) laboratory-space vector-potential,
$v\!\left(\vec{\chi}\right)$ is the interaction potential between the
particles, and $u\!\left(\vec{R}\right)$ is the \com trapping potential (in
free space, $u\!=\!0$).  Finally, $\vec{\chi} = \vec{r}_1 - \vec{r}_2$ and
$\vec{R} = (m_1/M) \vec{r}_1 + (m_2/M) \vec{r}_2$, where $M\!=\!m_1+m_2$. 

For systems of interest here, $m_1\!\ll\!m_2$.
Introducing $\mu\!=\!m_1 m_2/M$ and neglecting correction terms of the order
$O\!\left(\mu/M\right)$ in the laser interaction, Eq. \eqref{eqn:h-lab}
simplifies to\cite{Sindelka06a}:
\begin{align}
  \hat{H}_{\rm CoM} & = \hat{H}_\chi + \hat{H}_R & \label{eqn:h-bo} \\
  \hat{H}_\chi & = \frac{1}{2\mu} \left(\hat{p}_\chi+\vec{A}\!\left(\vec{R}+\vec{\chi},t\right)\right)^2
                 + v\!\left(\vec{\chi}\right) & \label{eqn:h-chi} \\
  \hat{H}_R & = \frac{1}{2M} \hat{p}_R^2 + u\!\left(\vec{R}\right)\text{.} & \label{eqn:h-R}
\end{align}
We have verified that the terms omitted in Eq. \eqref{eqn:h-bo} do not affect
the results reported below\cite{supplementary}.

The appropriate choice of the trapping potential $u\!\left(\vec{R}\right)$ in
Eq.~\eqref{eqn:h-R} and the shape of the initial \com wavepacket are the key
ingredients of our treatment. The extent of the \com wavepacket should be on
the order of the thermal de~Broglie wavelength of the target gas. The trapping
potential should not significantly disturb the targeted observables on the time
scale of the simulation. We have verified that the parabolic trapping potential
used presently satisfies these requirements\cite{supplementary}.

The general-case treatment of Eq.~\eqref{eqn:h-bo}, which contains a
non-separable coupling term through
$\vec{A}\!\left(\vec{R}+\vec{\chi},t\right)$, remains a formidable numerical
task. For the short (sub-picosecond) and moderately-intense IR fields, the \com
displacements remain small compared to both the characteristic electron
excursion and the laser-field wavelength. We therefore seek solutions of the
time-dependent Schr\"odinger equation (TDSE) in the close-coupling form:
\begin{align}
 \Psi\!\left(\vec{\chi},\vec{R},t\right) & = 
      \sum_{n} \phi_n\!\left(\vec{\chi},t\right) \zeta_n\!\left(\vec{R}\right)
    & \label{eqn:ansatz}
\end{align}
(From now on, we will omit arguments of $\phi_n$, $\zeta_n$ and other
spatially- and time-dependent quantities, as long as their choice is
unambiguous.) In Eq. \eqref{eqn:ansatz}, functions $\zeta_n$ are
orthonormalized, time-independent eigenfunctions of $\hat{H}_R$ (Eq.
\eqref{eqn:h-R}) with eigenvalues $\epsilon_n$.
We assume that the potential $u\!\left(\vec{R}\right)$ in Eq.~\eqref{eqn:h-R}
is such that the set of the discrete solutions $\{\zeta_n\}$ is complete.

Substituting the Ansatz \eqref{eqn:ansatz} into the TDSE for the Hamiltonian
\eqref{eqn:h-bo} and projecting on each $\zeta_m$ on the left, we obtain:
\begin{align}
  \ii \frac{\partial}{\partial t} \phi_m & = 
        \left(\hat{h}+\epsilon_m\right) \phi_m + \sum_n \hat{h}_{mn} \phi_n. & \label{eqn:tdse}
\end{align}
The explicit form of the one-electron operators $\hat{h}$ and $\hat{h}_{mn}$ is 
given by the Eqs.~\eqref{eqn:tdse-h}--\eqref{eqn:tdse-kappa}\cite{supplementary}.

The system of coupled PDEs \eqref{eqn:tdse} can be propagated in time at a cost
comparable to that of a standard, fixed-nuclei electronic TDSE, provided that
the number of the nuclear-coordinate channels is not excessive. 
At the end of the pulse, the expectation of a \com observable $\hat{O}$,
conditional on the internal degree of freedom being described by a normalized
wavefunction $\phi_a\!\left(\chi\right)$, is given by:
\begin{align}
\left<\hat{O}\right>_a = & \sum_{mn} \left<\zeta_m|\hat{O}|\zeta_n\right> 
                               \left<\phi_m|\phi_a\right>\left<\phi_a|\phi_n\right>\text{.}
	     	   & \label{eqn:expectation-conditional}
\end{align}
Choosing $\hat{O}=\hat{p}_R$ and $\hat{O}=\hat{1}$ yields the expectation of the
momentum and the state population, respectively. The \com 
velocity of the atom in an internal state $\phi_a$ is then:
\begin{align}
  \rmv_a = & \frac{1}{M} \frac{\left<\hat{p}_R\right>_a}{\left<\hat{1}\right>_a}\text{.} & \label{eqn:state-velocity}
\end{align}
We emphasize that the quantity $\rmv_a$ is determined from the expectation values
calculated after the field vanishes. It does not depend on field gauge choice,
and defines a physical observable.

We solve Eq.~\eqref{eqn:tdse} for a 3-dimensional hydrogen atom ($\mu\!=\!1$,
$M\!=\!1836$), initially in the $1s$ electronic ground state, exposed to a Gaussian
pulse of beam waist $w_0=30236$\bohr, central frequency $\omega=0.057$\au ($\lambda
\approx 799$~nm), and full-width-half-maximum $\tau_0=220$\au
($\approx5.32$~fs).  We choose for each Cartesian direction the following
convention: $x$--beam propagation, $y$--transverse, and $z$--polarization.  For
further details of the numerical parameters see\cite{supplementary}.

In a spatially non-uniform laser field, the excited atoms acquire the velocity
both in the forward and in the transverse directions. The final \com velocity
along laser polarization remains negligible, as required by symmetry.  We have
verified numerically that the forward velocity is insensitive to moderate
spatial-intensity gradients. As a result, we discuss the two components of the
velocity independently.

The forward (propagation-direction) component of velocity is a consequence of
the radiation pressure.  Strong-field excitation between hydrogenic levels with
the principal quantum numbers $n$ and $n'$ transfers the energy of $\Delta
E=0.5(n^{-2}-n'^{-2})$ from the laser field to the atom.  The corresponding
momentum transfer is $\Delta E/c$, giving the forward velocity:
\begin{align}
  \Delta \rmv_{\rm f} = & \frac{\Delta E}{M c}\text{.} & \label{eqn:vfwd}
\end{align}
Because it is determined solely by the initial and the final internal state of
the atom, it contains no information on the intervening dynamics. Our numerical
results (See Figs.~\ref{fig:beamcentre}, \ref{fig:w0half}\cite{supplementary})
are consistent with these expectations.

In the transverse direction the atoms are accelerated by the spatial gradient
of the ponderomotive potential.  Classically, the final outward velocity of an
initially-stationary particle with dipole polarizability $\alpha$ entering the
field at time $t_b$ in the vicinity of the beam waist 
($x\!=\!0$, Eq.~\eqref{eqn:Az}) is given by\cite{supplementary}:
\begin{align}
  \Delta \rmv_{\rm t} = & \frac{\alpha}{4 M} \int_{t_b}^{\infty} 
                       \frac{\partial}{\partial r}F_0^2\!\left(r,t\right) d t
                       \label{eqn:vtrans}
\end{align}
where $F_0\!\left(r,t\right)$ is the envelope of the 
laser electric field (see Eq.~\eqref{eqn:F0}). The hydrogen ground state
($\alpha_0=4.5$) is expected to be accelerated towards stronger fields
($\frac{\partial}{\partial r}F_0^2<0$).  Conversely, high-Rydberg states, which
exhibit the free-electron-like dynamical polarizabilities in low-frequency
fields ($\alpha_{\rm f}\!\approx\!-\omega^{-2}\approx-308$ at $799$~nm), are
expected to move towards weaker fields. 

A comparison of the calculated transverse velocity
(Eq.~\eqref{eqn:state-velocity}) with the classical Eq.~\eqref{eqn:vtrans} for
a state a known polarizability allows us to infer $t_b$ --- the time this state
has entered the field\cite{supplementary}. The integrand in
Eq.~\eqref{eqn:vtrans} is negative, so that $t_b$ is a monotonic function of
$\Delta \rmv_{\rm f}$ and defines a clock. Because $\alpha_{\rm f}$, the
low-frequency dynamical polarizability of the Rydberg states, is a
cycle-averaged quantity\cite{supplementary}, the time resolution of this clock
is $\approx1/2$ of the laser-cycle duration ($\approx1.3$~fs at 799~nm).


The composition of the Rydberg states populated by strong-field excitation is
sensitively affected by channel closings\cite{ZPI:17,PMM:17,Li14a}. We
therefore expect a similar effect to arise in the \com velocity spectroscopy.
At $799$~nm, channel closings occur each $26$~TW~cm$^{-2}$ ($\Delta I_{\rm
channel}\!=\!4\omega^3$). For a tightly-focused beam used presently
($w_0=2\lambda$), in the vicinity of the beam half-waist a channel closing
occurs each $648$\bohr, or $\approx\!34$~nm. We consider the channel-closing
effects by repeating the calculations at seven, equidistant transverse
points spaced by $216$\bohr, placed around the beam half-waist. We average the
results equally among these points.  This volume averaging effectively
suppresses resonance contributions, which are highly sensitive to the intensity
(See \cite{supplementary}).

\begin{figure}[!tbhp]
  \begin{center}
    \includegraphics[trim=0 0.7cm 0 0,width=\figurewidth]{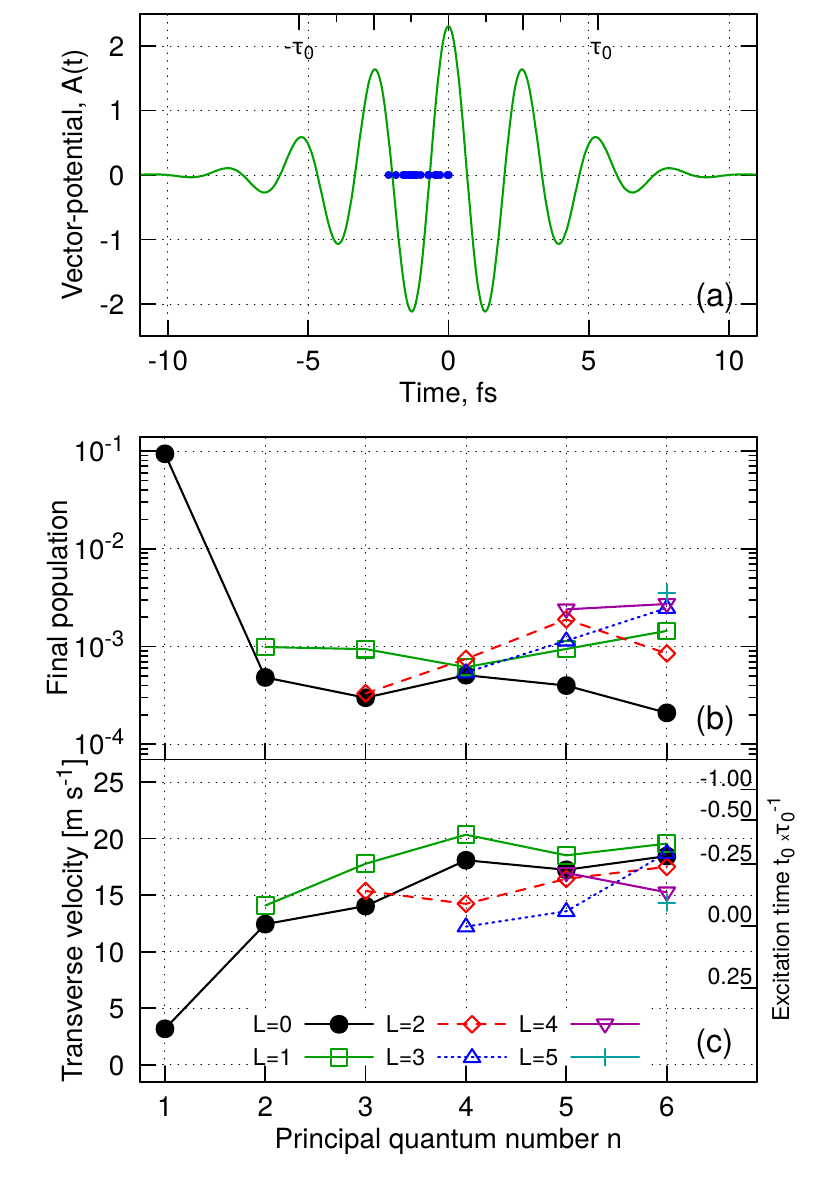} 
  \end{center}
  \caption{
(Color online) Hydrogen atom initially at the half-waist position. The results
are volume-averaged about the Cartesian point $(0,w_0/2\pm648,0)$.  The local
peak intensity is $\approx\!607$~TW~cm$^{-2}$. (a) Vector-potential at the
initial position as a function of time. The upper horizontal axis gives the
fraction of the pulse duration $\tau_0$. The blue dots on the time axis
indicate the reconstructed excitation times, see Fig.~\ref{fig:w0halfave:times}
for details. (b) Population of the individual $m\!=\!0$ bound states after the
end of the pulse.  (c) Final \com velocity in the outward transverse direction
in meters per second (1 atomic unit $\!\approx 2.19\!\times\!10^6$~m~s$^{-1}$).
The right vertical axis gives the time when a particle with $\alpha=\alpha_f$
needs to enter the field to reach the observed transverse velocity
(Eq.~\eqref{eqn:vtrans}).  The connecting lines in panels (b,c) are only a
guide for the eye.
} \label{fig:w0halfave}
\end{figure}

The maximum gradient of the ponderomotive potential occurs in the focal plane,
$w_0/2$ away from the focal spot.  We choose the point displaced in the $y$
direction, perpendicular to both the propagation and polarization directions.
The volume-averaged numerical results at this point are illustrated in
Fig.~\ref{fig:w0halfave}.  The local peak intensity of the field is
$\approx\!607$~TW~cm$^{-2}$.  The ionization is in the saturation regime, with
$\approx\!9\%$ of the population surviving in the $1s$ ground state after the
pulse.  Additionally, $\approx\!2.4\%$ of the atoms are excited to Rydberg
states with $n\!\le\!6$.  Although our simulation volume does not allow an
accurate determination of excitation probabilities for higher Rydberg states,
we estimate that at least $2\%$ of the atoms are left in Rydberg states with
$n\!\ge\!7$.  Most of the excited states possess magnetic quantum number
$m\!=\!0$, same as the initial state. 

\begin{figure}[!tbhp]
  \begin{center}
    \includegraphics[trim=0 0.7cm 0 0,width=\figurewidth]{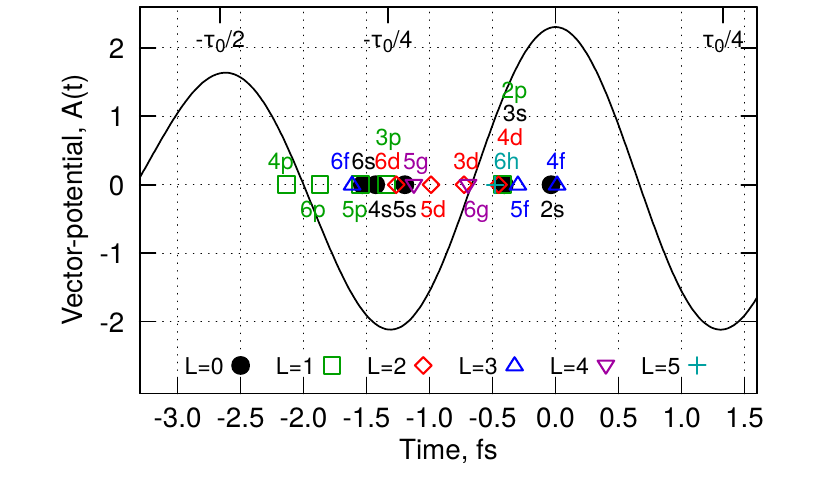} 
  \end{center}
  \caption{
(Color online) Reconstructed excitation times (See text and
Fig.~\ref{fig:w0halfave} for the raw data).  The vector potential at the
Cartesian point $(0,w_0/2,0)$ is given by the black solid line.  Peak of the
envelope is at the time zero. Please note that the resolution of the envelope
clock is $\approx1/2$ laser cycle ($\approx1.3$~fs).
} \label{fig:w0halfave:times}
\end{figure}

For all electronic states in Fig.~\ref{fig:w0halfave}c other than the ground
state, the final transverse velocities are in the range of
$12$--$20$~m~s$^{-1}$.  Solving Eq.~\eqref{eqn:vtrans} for $t_b$ yields the
excitation time. The results for the volume-averaged excitation time
reconstruction are presented in Fig.~\ref{fig:w0halfave:times}.  In all cases,
excited states are formed within the laser cycle immediately preceeding the
peak of the envelope. Although the excitation clock defined by the
Eq.~\eqref{eqn:vtrans} does not offer true sub-cycle resolution, it appears
that the Rydberg states with low principal quantum numbers tend to be populated
later in the laser pulse. This observation is consistent with the expectations
of the frustrated tunneling model\cite{NGS:08}: formation of the more compact,
low-$n$ states requires a tunnel exit point closer to the nucleus and
consequently higher electric field, reached closer to the peak of the envelope.

We present further fixed-intensity results
(Figs.~\ref{fig:beamcentre}--\ref{fig:w0half:times}), and explore the effects
of the carrier-envelope phase (CEP, Figs.~\ref{fig:w0halfavepihalf},
\ref{fig:w0halfavepihalf:times}), pulse duration
(Figs.~\ref{fig:w0halfavelong}, \ref{fig:w0halfavelong:times}), and
non-paraxial effects arising in a tightly-focused beam
(Figs.~\ref{fig:w0halfavepol}, \ref{fig:w0halfavepol:times}) in
\cite{supplementary}.  In all cases, we can successfully assign the preferred
excitation times based on the volume-averaged \com-velocity spectra, confirming
that the technique is universally applicable and experimentally realizable.
With a few exceptions, the reconstructed excitation times are before the peak
of the envelope, and tend to fall within the same laser cycle. For longer
pulses (See Figs.~\ref{fig:w0halfavelong},\ref{fig:w0halfavelong:times}), the
preferred excitation times shift to earlier times, before the peak of the
envelope. They however remain clustered within one laser cycle.

Because the ponderomotive clock is not sub-cycle accurate, we cannot
associate the time of the excitation with the specific phase of the
field. It may be possible to improve the time resolution of the excitation
clock using multi-color techniques, which have been successful for the
reconstruction of the ionization and recollision times in high-harmonic
spectroscopy\cite{Shafir12a,Bruner15a}.  Another possibility involves breaking
the symmetry of the interaction with a static, external magnetic field. Both
possibilities are currently under investigation.

\begin{figure}[!tbhp]
  \begin{center}
    \includegraphics[trim=0 0.7cm 0 0,width=\figurewidth]{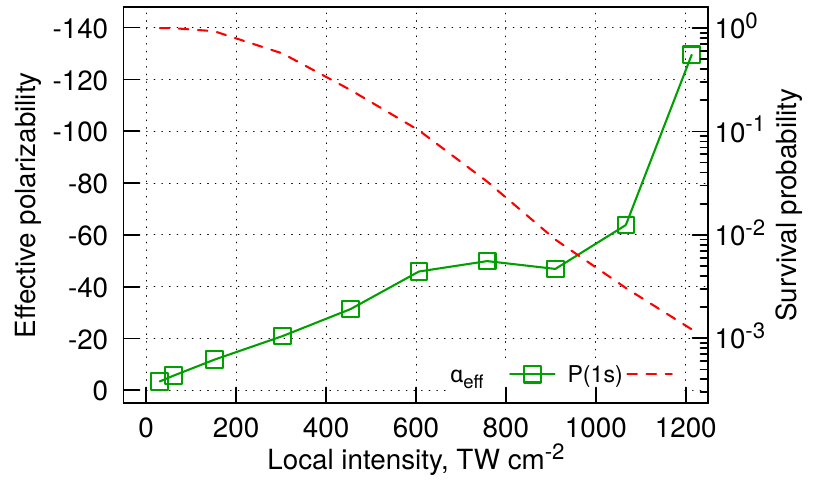} 
  \end{center}
  \caption{
(Color online) Effective polarizability $\alpha_{\rm eff}$ (green solid line;
left vertical axis) and survival probability (red
dashed line; right vertical axis) of the $1s$ ground state.  The
spatio-temporal field profile is the same as in Fig.~\ref{fig:w0halfave}. The
peak intensity $I_0$ varies from $50$~TW~cm$^{-2}$ to $2$~PW~cm$^{-2}$. The
horizontal axis shows the local peak intensity at the initial, half-waist
position of the atom ($I_{loc}\!\approx\!0.607\!\times\! I_0$).
} \label{fig:ground-polar}
\end{figure}

One remarkable result seen in Fig.~\ref{fig:w0halfave}c, which so far has not
been commented upon, is the behavior of the $1s$ ground state.  For the laser
pulse in Fig.~\ref{fig:w0halfave}a, it is \textit{weak-field} seeking, reaching
the final outward velocity of $\approx3.2$~m~s$^{-1}$.  The low-field-seeking
behavior of the $1s$ state persists for other field parameters as
well\cite{supplementary}.  The final $1s$ velocity is insensitive to
channel-closing effects, indicating that it arises due to adiabatic
modification of the ground state, rather than transient population of
high-Rydberg states. 

For the initial $1s$ state, $t_b\rightarrow-\infty$, and Eq.~\eqref{eqn:vtrans}
yields the effective polarizability $\alpha_{\rm eff}$, shown as a function of
the peak intensity of the laser pulse in Fig.~\ref{fig:ground-polar}. At
intensities below $50$~TW~cm$^{-2}$, the numerical accuracy is insufficient to
determine the final \com velocity ( Fig.~\ref{fig:velocity-1s}
\cite{supplementary}). The effective polarizability is \text{negative}, as
opposed to $+4.5$ expected for $1s$ in a weak field.  It is characteristic of
entering the Kramers-Henneberger regime\cite{WWK:17}.  Observation of
Kramers-Henneberger regime for an atomic ground state in strong, low-frequency
fields has been long sought after, with no unambiguous detection thus
far\cite{WWK:17}.

To summarize, we have developed a computationally-tractable quantum mechanical
approach to correlations between \com motion and internal electronic dynamics
in strong, non-uniform laser fields. Using the technique, we demonstrate that
the final \com velocity is sensitive to the internal excitation dynamics.  In
particular the transverse, ponderomotive velocity is determined by the total
time the excited state spends in the field. In the absence of resonances, it
yields a measurement of the preferential time of excitation.  This procedure is
robust to limited volume averaging, and can be applied for different CEP
values, for longer pulses, and for non-paraxial beams.  Finally, we demonstrate
an unambiguous signature of the atomic ground state entering the
Kramers-Henneberger regime in strong, low-frequency fields, which has been long
sought-for. Taken together, our results suggest that \com-velocity
spectroscopy is a powerful, and so far overlooked tool for understanding
strong-field bound-state electronic dynamics on their natural timescale.

We expect that similar ideas, using a collective, nearly-classical degrees of
freedom of a quantum system as an intrinsic measurement device may become
useful in other contexts as well.
\nocite{Manolopoulos02a,Lax75a,Patchkovskii16a}

\bibliography{bib_ei}
\makeatletter\@input{xx-support.tex}\makeatother

\end{document}


\title{Dissecting strong-field excitation dynamics with atomic-momentum spectroscopy}

\author{A.W.\ Bray}
\email{alexander.bray@anu.edu.au}
\affiliation{Australian National University, Canberra}
\author{U.\ Eichmann}
\email{eichmann@mbi-berlin.de}
\author{S.\ Patchkovskii}
\email{serguei.patchkovskii@mbi-berlin.de}
\affiliation{Max-Born-Institute, Berlin, Germany} 

\date{\today}

\maketitle

\section{Supplemental Material}

\renewcommand\thefigure{S\arabic{figure}}
\renewcommand\theequation{S\arabic{equation}}
\setcounter{figure}{0}
\setcounter{equation}{0}

\subsection{Choice of the trapping potential and the \com wavepacket}

Unless noted otherwise, we apply the parabolic trapping potential
$u\!\left(\vec{R}\right)\!\!=\!\!k R^2/2$ with $k\!\!=\!\!10^{-4}$. We include
four \com eigenstates: the initially-populated ground state and the three
singly-excited harmonic vibrational modes, one along each Cartesian direction.
For this choice of the force constant $k$, the free-oscillation period in the
parabolic trap is $\approx650$~fs, much longer than any of the pulses
considered in the simulations.  For simulations discussed below and in the main
text, electronic populations associated with the excited \com states do not
exceed $7\!\times\!10^{-3}$. The populations of the higher trap modes are
therefore expected to be $<10^{-4}<\left(7\!\times\!10^{-3}\right)^2$ and can
be safely neglected for our purposes.

The spatial extent (the full width at the half-maximum) of the initial Gaussian
wavepacket in $\approx2.6a_0$. This value should be compared to the thermal
de~Broglie wavelength of the target gas $\Lambda$:
\begin{align}
  \Lambda &= \left(\frac{2\pi}{M k T}\right)^{1/2}. & \label{eqn:deBroglie}
\end{align}
For a hydrogen atom ($M\approx1837$), $\Lambda=2.6$ corresponds to the target
gas temperature of $\approx160$~K. This value is consistent with the temperature
expected for an effusive source.

We have repeated some of the simulations using a modified trap with
$k'\!=\!3k\!=\!3\!\times\!10^{-4}$.  This choice decreases the final
populations of the excited \com modes by a factor of $\sqrt{k/k'}=1/\sqrt{3}$.
The final, state-resolved \com velocities remain numerically unchanged,
confirming that the trap does not significantly influence the \com dynamics on
the time scale of the simulation.

\subsection{Explicit form of the Eq.~\eqref{eqn:tdse}}

Our working equations are given by Eq.~\eqref{eqn:tdse} of the main text:
\begin{align}
  \ii \frac{\partial}{\partial t} \phi_m & = 
        \left(\hat{h}+\epsilon_m\right) \phi_m + \sum_n \hat{h}_{mn} \phi_n & \nonumber 
\end{align}
The operators $\hat{h}$ and $\hat{h}_{mn}$ are given by:
\begin{align}
  \hat{h} & = \frac{1}{2\mu}\hat{p}_\chi^2 + v 
            + \frac{1}{\mu} \vec{A}\cdot\hat{p}_\chi + \frac{1}{2\mu}A^2 & \label{eqn:tdse-h} \\
  \hat{h}_{mn} & = \vec{\eta}_{mn}\cdot\hat{p}_\chi + \kappa_{mn} & \label{eqn:tdse-hmn} \\
  \vec{\eta}_{mn} & = \frac{1}{\mu} \sum_{b=x,y,z} \vec{A}^{(b)} 
                   \left<\zeta_m|R_b|\zeta_n\right> & \label{eqn:tdse-eta} \\
  \kappa_{mn} & = \frac{1}{2\mu} \sum_{b=x,y,z} \left(\frac{\partial}{\partial\chi_b}A^2\right)
                    \left<\zeta_m|R_b|\zeta_n\right> \nonumber \\
         & + \frac{1}{2\mu} \sum_{b,c=x,y,z} \vec{A}^{(b)}\cdot\vec{A}^{(c)}
                    \left<\zeta_m|R_b R_c|\zeta_n\right>\text{.} & \label{eqn:tdse-kappa}
\end{align}
In Eqs. \eqref{eqn:tdse-h}--\eqref{eqn:tdse-kappa}, $\vec{A}$ and $\vec{A}^{(b)}$ are 
functions of the interparticle distance $\vec{\chi}$ and time $t$. We further assume
that the higher-order terms in the Taylor expansion for the \com coordinate:
\begin{align}
  \vec{A}\!\left(\vec{R}+\vec{\chi},t\right) & = 
                      \vec{A}\!      \left(\vec{\chi},t\right)
     + \!\!\!\!\sum_{a=x,y,z} \!\!\vec{A}^{(a)}\!\left(\vec{\chi},t\right) R_a
     + \ldots & \label{eqn:taylor-a}
\end{align}
can be neglected. This assumption is valid for non-relativistic IR fields and
short pulse durations, where nuclear displacements remain small compared to the
laser wavelength.

\subsection{Treatment for the general masses and charges}

The minimal-coupling non-relativistic Hamiltonian for two particles is given
by:
\begin{align}
  \hat{H} & = \frac{1}{2m_1} \left(\hat{p}_1 - q_1 \vec{A}\!\left(\vec{r}_1,t\right)\right)^2
            + \frac{1}{2m_2} \left(\hat{p}_2 - q_2 \vec{A}\!\left(\vec{r}_2,t\right)\right)^2 \nonumber \\
          & + v\!\left(\vec{\chi}\right) + u\!\left(\vec{R}\right) & \label{eqn:h-lab:general}
\end{align}
where $m_i$, $q_i$, and $\hat{p}_i$, $i\!=\!1,2$ are respectively the mass, charge,
and momentum operator for particles $1$ and $2$. Using the standard \com
variable substitution (see the main text), we obtain, for the transverse field
$\vec{A}$:
\begin{align}
  \hat{H} & = \frac{1}{2\mu}\hat{p}_\chi^2 
            - \left(\frac{1}{m_1}\vec{A}_1-\frac{1}{m_2}\vec{A}_2\right)\cdot\hat{p}_\chi
            + \frac{1}{2m_1}A_1^2 \nonumber \\ &
            + \frac{1}{2m_2}A_2^2
            + v\!\left(\vec{\chi}\right)
            + \frac{1}{2M}\hat{p}_R^2
            - \frac{1}{M}\left(\vec{A}_1+\vec{A}_2\right)\cdot\hat{p}_R \nonumber \\ &
            + u\!\left(\vec{R}\right)
            - \frac{1}{\mu} R_a \left(\vec{A}_1^{(a)}+\vec{A}_2^{(a)}\right)\cdot\hat{p}_\chi \nonumber \\ &
            - R_a\left(\frac{1}{m_2}\vec{A}_1^{(a)}-\frac{1}{m_1}\vec{A}_2^{(a)}\right)\cdot\hat{p}_R \nonumber \\ &
            + \frac{1}{\mu} R_a \left(\vec{A}_1\cdot\vec{A}_1^{(a)}-\vec{A}_2\cdot\vec{A}_2^{(a)}\right) \nonumber \\ &
            + \frac{1}{2\mu} R_a R_b \left(\frac{M}{m_2}\vec{A}_1^{(a)}\cdot\vec{A}_1^{(b)}
                                          +\frac{M}{m_1}\vec{A}_2^{(a)}\cdot\vec{A}_2^{(b)}\right) &
   \label{eqn:h-nonbo}
\end{align}
where summation over indices $a$ and $b$ is implied, and quantities $\vec{A}_i$
and $\vec{A}_i^{a}$ ($i\!=\!1,2$; $a\!=\!x,y,z$) are given by:
\begin{align}
  \vec{A}_1\!\left(\vec{\chi}\right) & = q_1 \vec{A}\!\left( \frac{m_2}{M}\vec{\chi}\right) & \label{eqn:a1} \\
  \vec{A}_2\!\left(\vec{\chi}\right) & = q_2 \vec{A}\!\left(-\frac{m_1}{M}\vec{\chi}\right) & \label{eqn:a2} \\
  \vec{A}_i^{(a)}\!\left(\vec{\chi}\right) & = 
          \frac{\partial}{\partial\chi_a}\vec{A}_i\!\left(\vec{\chi}\right)\text{.} & \label{eqn:ad}
\end{align}
In deriving Eq.~\eqref{eqn:h-nonbo} we assumed that $\vec{R}$ is small compared
to the laser wavelength, so that the higher-order terms can be neglected in
Eq.~\eqref{eqn:taylor-a}.

Equation~\eqref{eqn:h-nonbo} has been implemented numerically within the Ansatz
of the Eq.~\eqref{eqn:ansatz}, and is propagated numerically for the same grids
and trap parameters as the Eq.~\eqref{eqn:tdse} in the main text. The
propagation cost is $\approx\!\!5\!\times$ that of Eq.~\eqref{eqn:tdse}.  For
atomic hydrogen, the results on the scale of the graphs shown in the main text
and the supplementary material are indistinguishable from those of the
approximate Eq.~\eqref{eqn:tdse}.

\subsection{Discretization and numerical integration}

The electronic wavefunction is discretized on a uniform,
Cartesian-product grid extending to $\pm78.6$\bohr in the $X$ and $Y$
directions, and to $\pm152.1$\bohr along $Z$ (the laser polarization
direction).  Uniform grid spacing is $0.35$\bohr.  A transmission-free absorbing
boundary\cite{Manolopoulos02a} is applied starting $9.4$\bohr from all grid edges.

Eqs. \eqref{eqn:tdse} are integrated using the leap-frog propagator with the
time step of $0.005$\au.  Using the 4th-order Runge-Kutta propagator leads to
numerically equivalent results. We analyze the results by projecting the final
wavefunctions onto hydrogenic states (Eq.
\eqref{eqn:expectation-conditional}).  Our simulation volume is sufficient to
resolve Rydberg states up to the principal quantum number $n\!=\!6$. We
verified that a further increase in the simulation volume does not affect these
results.

\subsection{Beam profile}

The laser beam propagates toward the positive $X$ direction. We model the
vector-potential of the beam by a paraxial TEM$_{00}$ mode, including the
lowest-order longitudinal correction\cite{Lax75a}:
\begin{align}
  A_z &= A_0 f\left(t-\frac{x}{c}\right) \Re\left[\left(1-\frac{\ii x}{z_R}\right)^{-1}
              \E^\Phi \right] & \label{eqn:Az} \\
  A_x &= A_0 f\left(t-\frac{x}{c}\right) \Re\left[\frac{\ii z}{z_R}\left(1-\frac{\ii x}{z_R}\right)^{-2}
              \E^\Phi \right] & \label{eqn:Ax} \\
  \Phi &= -\ii k x + \ii \omega t + \ii \varphi_0 - \frac{r^2}{w_0^2}
              \left(1-\frac{\ii x}{z_R}\right)^{-1} & \label{eqn:Phi}
\end{align}
where the beam waist $w_0=30236$\bohr, Rayleigh range $z_R = \pi w_0/\lambda$,
$\lambda = 2\pi c/\omega$ is the wavelength, $k=2\pi/\lambda$ is the
wavevector, and $c\approx137$ is the speed of light. Unless noted otherwise,
the slowly-varying envelope $f(t-x/c)$ is a truncated
Gaussian\cite{Patchkovskii16a} with the FWHM of $220$\au ($\approx
5.32$~fs) and baseline duration of $800$\au ($\approx\!19.4$~fs). The
carrier frequency $\omega\!=\!0.057$, corresponding to $\lambda\!\approx\!799$~nm.
The peak vector-potential at the focal-spot center is $2.9615$, corresponding
to a peak field intensity of $\approx\!1$~PW~cm$^{-2}$.  

In the vicinity of the beam waist ($x\!=\!0$), the transverse component of the
vector-potential becomes:
\begin{align}
 A_z &\approx \frac{F_0}{\omega} \cos\left(\omega t - k x + \phi_0\right), & \label{eqn:Az-waist}
\end{align}
where
\begin{align}
 F_0 &= \omega A_0 f\left(t-\frac{x}{c}\right) \E^{-r^2/w_0^2}. & \label{eqn:F0}
\end{align}
For a slowly-varying envelope $f$, the laser electric field is given by:
\begin{align}
 F_z &= -\frac{\partial}{\partial t}A_z \approx F_0 \sin\left(\omega t - k x + \phi_0\right)
     & \label{eqn:Fz}
\end{align}
All numerical calculations reported in the main text and below used the full
vector-potential expressions (Eqs.~\eqref{eqn:Az}-\eqref{eqn:Phi}). The 
envelope approximation of Eq.~\eqref{eqn:F0} is however extremely useful 
for the classical interpretation of the results (See Eq.~\eqref{eqn:vtrans}
in the main text and its derivation below).

\subsection{Derivation of the Eq.~\eqref{eqn:vtrans}}

Let us consider the motion of a classical particle with charge $\cq$, mass $\cm$,
and dipole polarizability $\alpha_0$. The particle enters the field of a laser
pulse at the time $t_b$, at the position $\vec{\rmr}_b$, with the velocity
$\vec{\rmv}_b$. We assume that the displacement of the particle remains small
compared to the wavelength of the laser field and to the characteristic scale
of spatial inhomogeneity.  Under these assumptions we can treat the spatial
variation of the laser field as a perturbation to the motion in the uniform
field.

Newton's equation of motion for the particle in a uniform field
$F\!\left(t\right)$ are given by:
\begin{align}
 \overset{..}{\vec{\rmr}} &= \frac{\cq}{\cm} \vec{F}\left(t\right) & \label{eqn:classical:a}
\end{align}
Integrating over the time, we obtain:
\begin{align}
 \overset{.}{\vec{\rmr}} &= 
                \vec{\rmv}_b + \frac{\cq}{\cm}\int_{t_b}^{t} \vec{F}\left(t'\right) dt' & \nonumber \\
             &= \vec{\pi} - \frac{\cq}{\cm} \vec{A}\left(t\right), & \label{eqn:classical:v}
\end{align}
where
\begin{align}
  \vec{A}\left(t\right) &= -\int_{-\infty}^{t} \vec{F}\left(t'\right) dt' & \label{eqn:classical:vp}
\end{align}
is the vector-potential of the field, and
\begin{align}
  \vec{\pi} &= \vec{\rmv}_b + \frac{\cq}{\cm}\vec{A}\left(t_b\right) & \label{eqn:classical:pi}
\end{align}
is the drift velocity. Integrating the second time, we get:
\begin{align}
  \vec{\rmr} &= \vec{\rmr}_b + \vec{\pi} \left(t-t_b\right) 
              - \frac{\cq}{\cm} \int_{t_b}^{t} \vec{A}\left(t'\right) dt'. & \label{eqn:classical:r}
\end{align}

The total energy $E$ of the particle is then given by:
\begin{align}
  E &= E_0 - \frac{\alpha_0}{2} \vec{F}^2 + \frac{\cm}{2} {\overset{.}{\vec{r}}}^2 & \nonumber \\
    &= E_0 - \frac{\alpha_0}{2} \vec{F}^2 + \frac{\cm}{2} \vec{\pi}^2 + \frac{\cq^2}{2\cm}\vec{A}^2
           - \cq \vec{\pi}\cdot\vec{A}. & \label{eqn:classical:etot}
\end{align}
In the slowly-varying envelope approximation (cf. Eqs.~\eqref{eqn:Az-waist},\eqref{eqn:Fz}),
\begin{align}
  \vec{A} &\approx \frac{1}{\omega}\vec{F}_0\left(r,t\right) 
           \Re{\E^{\ii\left(\omega t-\vec{k}\cdot\vec{r}+\phi_0\right)}} \label{eqn:classical:A} \\
  \vec{F} &\approx                 \vec{F}_0\left(r,t\right) 
           \Im{\E^{\ii\left(\omega t-\vec{k}\cdot\vec{r}+\phi_0\right)}} \label{eqn:classical:F}
\end{align}
where the envelope $\vec{F}_0\left(r,t\right)$ is a slowly-varying function of
both arguments (we omit $\vec{F}_0$ arguments from now on). Replacing all
rapidly-oscillating quantities in Eq.~\eqref{eqn:classical:etot} by their
carrier-cycle averages, we then obtain:
\begin{align}
 \left<E\right> &= E_0 + \frac{\cm}{2}\vec{\pi}^2 - \frac{\alpha_0}{4}\vec{F}_0^2
                 + \frac{\cq^2}{4\cm\omega^2} \vec{F}_0^2, & \label{eqn:classical:eave}
\end{align}
where $E_0$ is the energy of the particle in the absence of the field.

For a weakly-inhomogeneous field envelope $F_0$, the additional momentum imparted
on the particle is given by the perturbation-theory expression:
\begin{align}
 \Delta\vec{p} &= -\int_{t_b}^\infty \frac{\partial}{\partial\vec{r}}\left<E\right> dt & \nonumber \\
               &= \frac{\alpha}{4} \int_{t_b}^\infty \frac{\partial}{\partial\vec{r}}\vec{F}_0^2 dt,
                \label{eqn:classical:dp}
\end{align}
where
\begin{align}
 \alpha &= \alpha_0 - \frac{\cq^2}{\cm\omega^2} & \label{eqn:classical:alpha}
\end{align}
For the ground-state hydrogen atom in the weak-field limit, $\cq=0$ and
$\alpha=\alpha_0=4.5a_0^3$. For a free electron, or a bound state in 
the asymptotic Kramers-Henneberger regime, $\alpha_0=0$, $\cq=-1$, $\cm=1$,
and $\alpha=\alpha_f=-\omega^{-2}$.

For states which remain bound after the end of the pulse, the drift velocity of
the electron $\vec{\pi}\approx0$, and the entire momentum $\Delta\vec{p}$ is
transferred to the center of mass of the atom. Dividing $\Delta\vec{p}$ of the
Eq.~\eqref{eqn:classical:dp} by the total mass $M$ we therefore obtain
Eq.~\eqref{eqn:vtrans} of the main text. 

From the cycle-average derivation above, it is clear that
Eq.~\eqref{eqn:vtrans} does not provide sub-cycle resolution for $t_b$.
Rather, the clock defined by Eq.~\eqref{eqn:vtrans} corresponds to the time
under the pulse envelope.

Substituting the specific envelope \eqref{eqn:F0} used in our numerical
simulations into Eq.~\ref{eqn:vtrans}, and limiting the consideration to the
beam-waist plane, we obtain:
\begin{align}
  \Delta \rmv_{\rm t} &= -\frac{\alpha}{4M} 
                      \frac{\omega^2A_0^2\tau_0 r}{w_0^2} \sqrt{\frac{\pi}{\ln{2}}} 
                      \E^{-2r^2/w_0^2}
                      {\rm erfc}\!\left(2\sqrt{\ln{2}}\frac{t_b}{\tau_0}\right),
  \label{eqn:vtrans:explicit}
\end{align}
where $\tau_0$ is the FWHM duration of the laser pulse.
Eq.~\ref{eqn:vtrans:explicit} is a monotonous function of $t_b$. Solving it for
$t_b$ with $\alpha=-\omega^{-2}$ yields the desired envelope clock.

\subsection{Additional results}

\begin{figure}[!tbhp]
  \begin{center}
    \includegraphics[trim=0 0.7cm 0 0,width=\figurewidth]{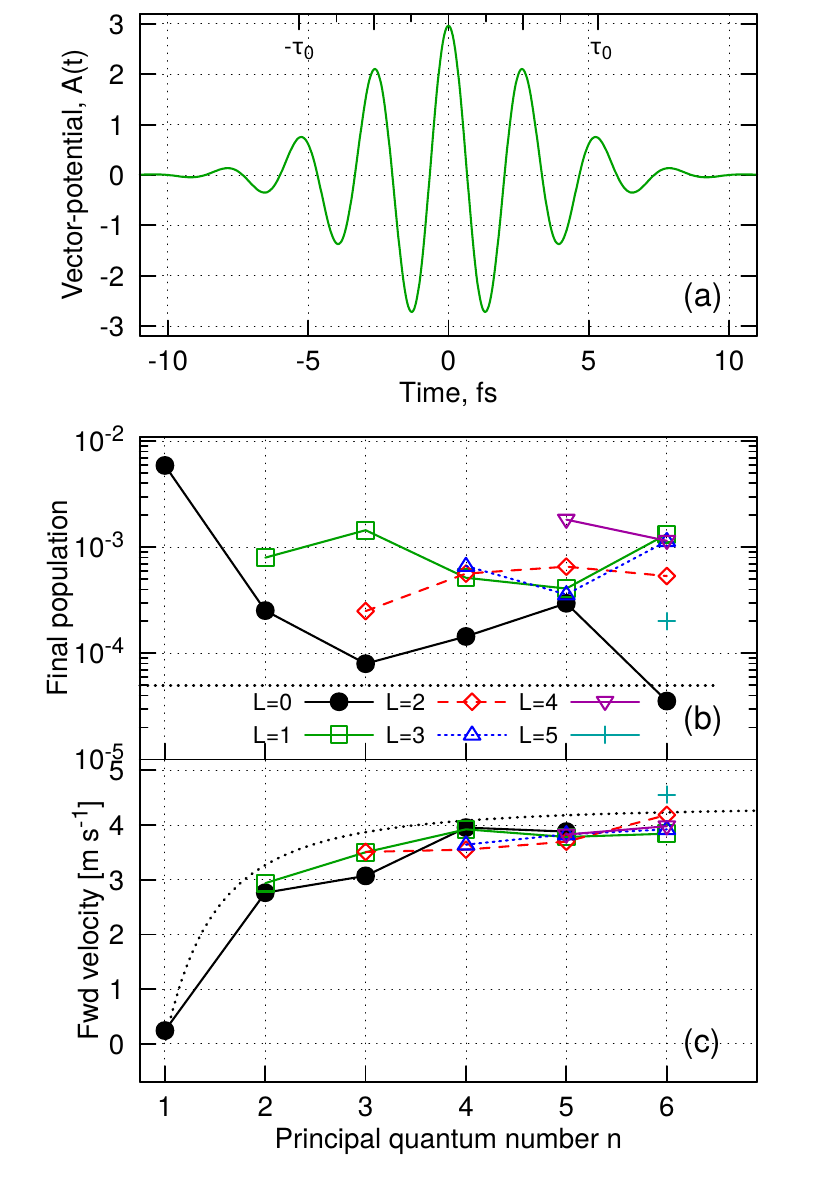} 
  \end{center}
  \caption{
(Color online) Hydrogen atom initially at the center of the focal spot. Local
peak intensity 1~PW~cm$^{-2}$. Also see Fig.~\ref{fig:w0half} caption.
} \label{fig:beamcentre}
\end{figure}

We begin by briefly examining the dynamics of a hydrogen atom at the center of
the focal spot (Fig.~\ref{fig:beamcentre}), where the laser field remains
essentially a planewave. At the $1$~PW~cm$^{-2}$ local peak intensity, ionization
is in the saturation regime. Only $0.6\%$ of the hydrogen atoms remain in the
$1s$ ground state. Additionally, approximately $1.3\%$ of the atoms are excited
to Rydberg states with $n\!\le\!6$.  We further estimate that at least $2\%$ of
the atoms are left in Rydberg states with $n\!\ge\!7$.  Most of the excited
states possess magnetic quantum number $m\!=\!0$, same as the initial state.
Due to their low population, we do not consider $m\!\ne\!0$ states any further.
The individual excitation probabilities and final \com velocities are collected
in Fig.~\ref{fig:beamcentre}. These results are qualitatively similar to those
at the half-waist position, discussed below and in the main text.

The final populations of the $m\!=\!0$ states are illustrated in
Fig.~\ref{fig:beamcentre}b.  The corresponding \com velocities in the
propagation direction are shown in Fig.~\ref{fig:beamcentre}c. We find that the
calculated velocities become erratic for very low final populations, which
amplify the inevitable numerical noise in the solutions (see
Eq.~\eqref{eqn:state-velocity}). As a result, we choose to disregard the
calculated \com velocities when the corresponding population drops below
$5\!\times\!10^{-5}$ (e.g. for the $6s$ state in Fig.~\ref{fig:beamcentre}c).  The
net number of photons absorbed from the laser field is zero for the $1s$ ground
state, which consequently acquires no forward velocity. For all other states in
Fig.~\ref{fig:beamcentre}c, the final forward velocity is in the range of
$2.8$~m~s$^{-1}$ ($2s$) to $4.6$~m~s$^{-1}$ ($6h$).  These values are
consistent with the momentum transfer due to the radiation pressure
(Eq.~\eqref{eqn:vfwd}), and carry no information on the internal dynamics of
the system.

\begin{figure}[!tbhp]
  \begin{center}
    \includegraphics[trim=0 0.7cm 0 0,width=\figurewidth]{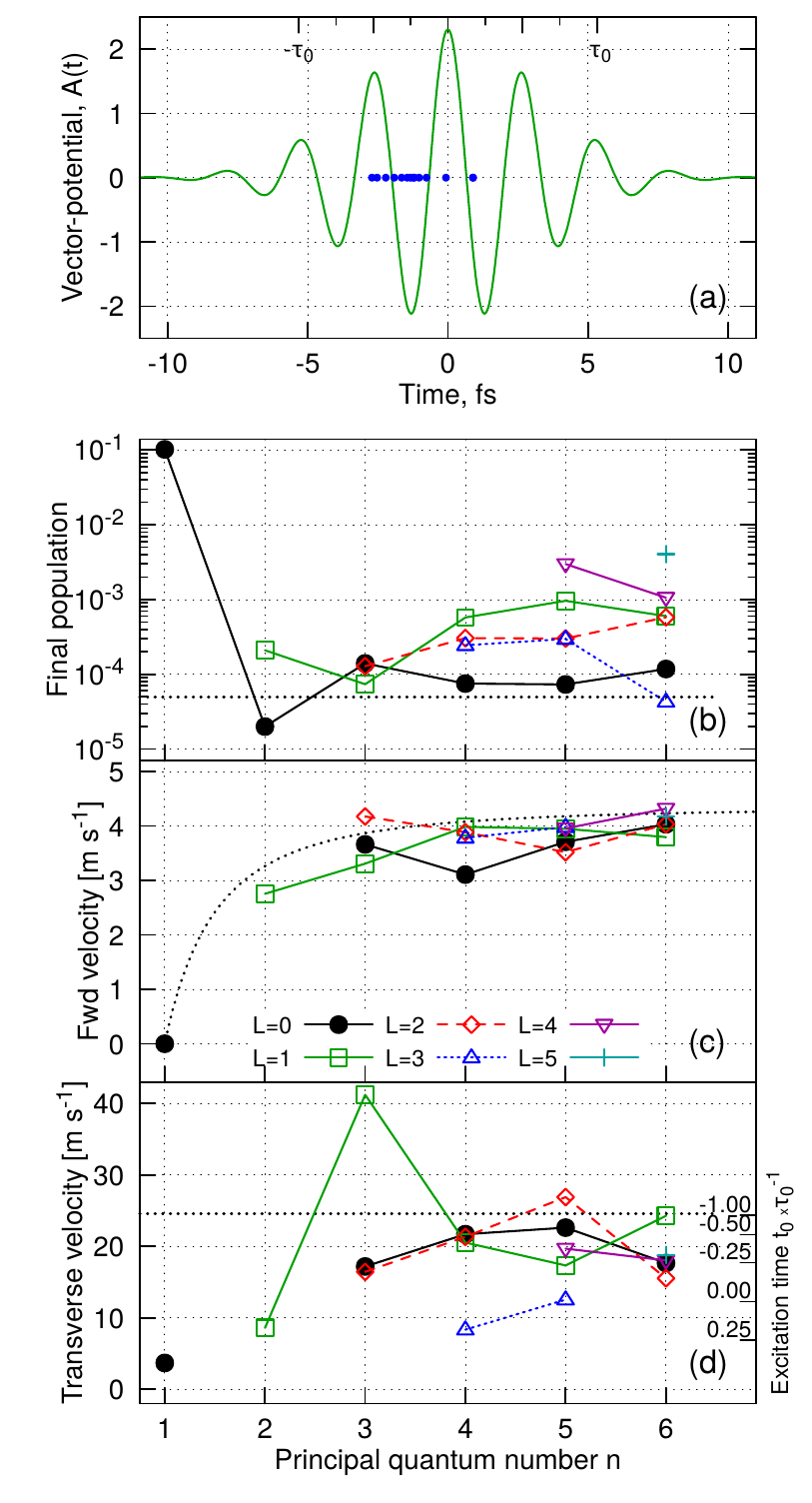} 
  \end{center}
  \caption{
(Color online) Hydrogen atom initially at the beam half-waist (the Cartesian
position $(0,w_0/2,0)$; no spatial averaging).  Local peak intensity
$\approx\!607$~TW~cm$^{-2}$. (a) Vector-potential at the initial position as a
function of time. The upper horizontal axis gives the fraction of the pulse
duration $\tau_0$. The blue dots on the time axis indicate the reconstructed
excitation times (see Fig.~\ref{fig:w0half:times}). (b) Population of the
individual $m\!=\!0$ bound states after the end of the pulse.  Horizontal
dotted line indicates the acceptance cut-off (see text).  (c) Final \com
velocity in the beam propagation direction in meters per second (1 atomic unit
$\!\approx\!2.19\!\times\!10^6$~m~s$^{-1}$).  The dotted line indicates the
$\Delta E/M c$ contribution to the final velocity expected from the radiation
pressure.  (d) Final \com velocity in the outward transverse
direction.  The right vertical axis gives the time when a particle with
$\alpha=\alpha_f$ needs to enter the field to reach the observed transverse
velocity (Eq.~\eqref{eqn:vtrans}). Final velocities above the dotted horizontal
line cannot be reached for $\alpha=\alpha_f$ (see text).  The connecting lines
in panels (b--d) are only a guide for the eye.
} \label{fig:w0half}
\end{figure}

\begin{figure}[!tbhp]
  \begin{center}
    \includegraphics[trim=0 0.7cm 0 0,width=\figurewidth]{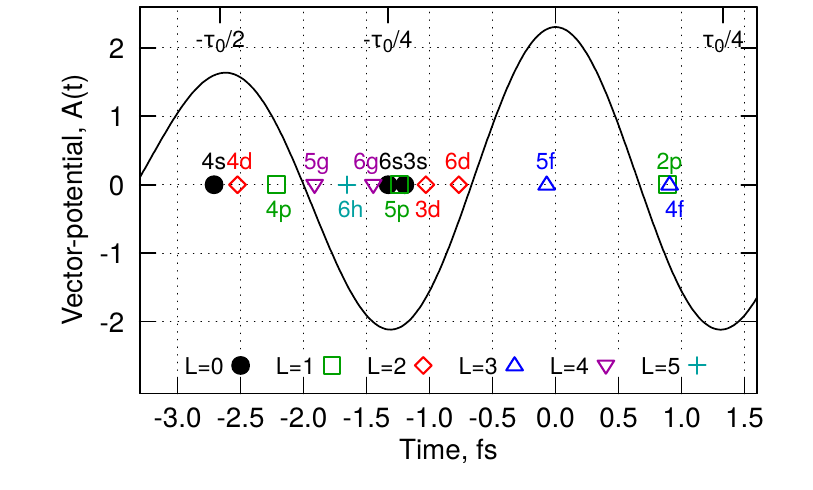} 
  \end{center}
  \caption{
(Color online) Reconstructed excitation times for the \com velocity spectra in
Fig.~\ref{fig:w0half} (see text). No excitation times can be assigned for the
$3p$, $5s$, $5d$, and $6p$ states, which are affected by resonances, or for the
$1s$ ground state. The solid line is the vector-potential from
Fig.~\ref{fig:w0half}. Note that the nominal resolution of the envelope clock
(Eq.~\eqref{eqn:vtrans}) is $\approx1/2$ of the carrier cycle
($\approx1.3$~fs).
} \label{fig:w0half:times}
\end{figure}

We now turn to positions away from the center of the focal spot, where the
laser field is no longer well-approximated by a planewave. As discussed in the
main text, the maximum of the spatial gradient of the ponderomotive potential
is found on a ring of the radius $w_0/2$ around the focal spot, normal to the
beam propagation direction.  We begin by choosing the point displaced in the
$y$ direction, perpendicular to both the propagation and polarization
directions. The non-paraxial corrections of Eq.~\eqref{eqn:Ax} vanish in the
vicinity of this point.  The local peak intensity of the field is
$\approx\!607$~TW~cm$^{-2}$.  The ionization remains in the saturation regime,
with $\approx\!10\%$ of the population surviving in the ground state after the
pulse. The numerical results at this point are illustrated in
Fig.~\ref{fig:w0half}. The final populations of the Rydberg states are of a
similar overall magnitude to those found at the beam center, with
$\approx\!1.4\%$ left in Rydberg states with $n\!\le\!6$, and $\ge\!3\%$ in states with
$n\!\ge\!7$.

The forward velocities at the half-waist position (Fig~\ref{fig:w0half}c)
follow a trend similar to the beam center, and are consistent with the
radiation-pressure effects. The forward velocity is insensitive to the
small transverse intensity gradient. Repeating the calculation with a
planewave field at the $607$~TW~cm$^{-2}$ peak intensity yields results
visually identical to Fig.~\ref{fig:w0half}c.

From Eq.~\eqref{eqn:vtrans}, we expect the transverse, ponderomotive
acceleration to yield the maximum velocity when a Rydberg state is formed early
within the laser pulse ($t_b\rightarrow-\infty$). This sets the upper bound on
the final transverse velocity a state with $\alpha=\alpha_f\approx-308$ could
reach. For the field parameters in Fig.~\ref{fig:w0half}, the limit is
$24.6$~m~s$^{-1}$, indicated by the dotted line in panel (d). This limit is
exceeded by the $3p$ and $5d$ states, indicating that the magnitude of their
effective polarizability exceeds the free-electron value.  In the case of the
$3p$ state, which reaches the final transverse velocity of $41$~m~s$^{-1}$, we
speculate that the likely reason is a 1-photon resonance with the Stark-shifted
$2s$ state ($1.9$~eV away in the absence of the field).  The resonance depleted
the $2s$ state (final population of $1.9\!\times\!10^{-5}$), while
simultaneously increasing the effective polarizability and the acceleration of
the $3p$ state. Because strong-field resonances are highly-sensitive to the
local peak intensity, we anticipate such resonances to be washed out by the
spatial averaging (see below and in the main text).

We can now invert Eq.~\eqref{eqn:vtrans} to extract the time estimate $t_b$. We
assume that the frequency-dependent polarizability of the final state is
$\alpha_f=-\omega_0^{-2}$ ($\approx-308$). Away from resonances, this
estimation becomes progressively more accurate for higher Rydberg states.  The
resulting excitation-time reconstruction is shown in
Fig.~\ref{fig:w0half:times}.

In the volume-averaged simulation (Fig.~\ref{fig:w0halfave} of the main text),
$t_b$ can be consistently assigned for all final states.  Because suppression
of resonances appears to be important for a successful reconstruction of the
excitation time, all subsequent results use volume-averaging protocol similar
to that in Fig.~\ref{fig:w0halfave} ($\pm648$\bohr along the maximum
intensity-gradient direction, covering $\approx\!\!2$ channel closings).

\begin{figure}[!tbhp]
  \begin{center}
    \includegraphics[trim=0 0.7cm 0 0,width=\figurewidth]{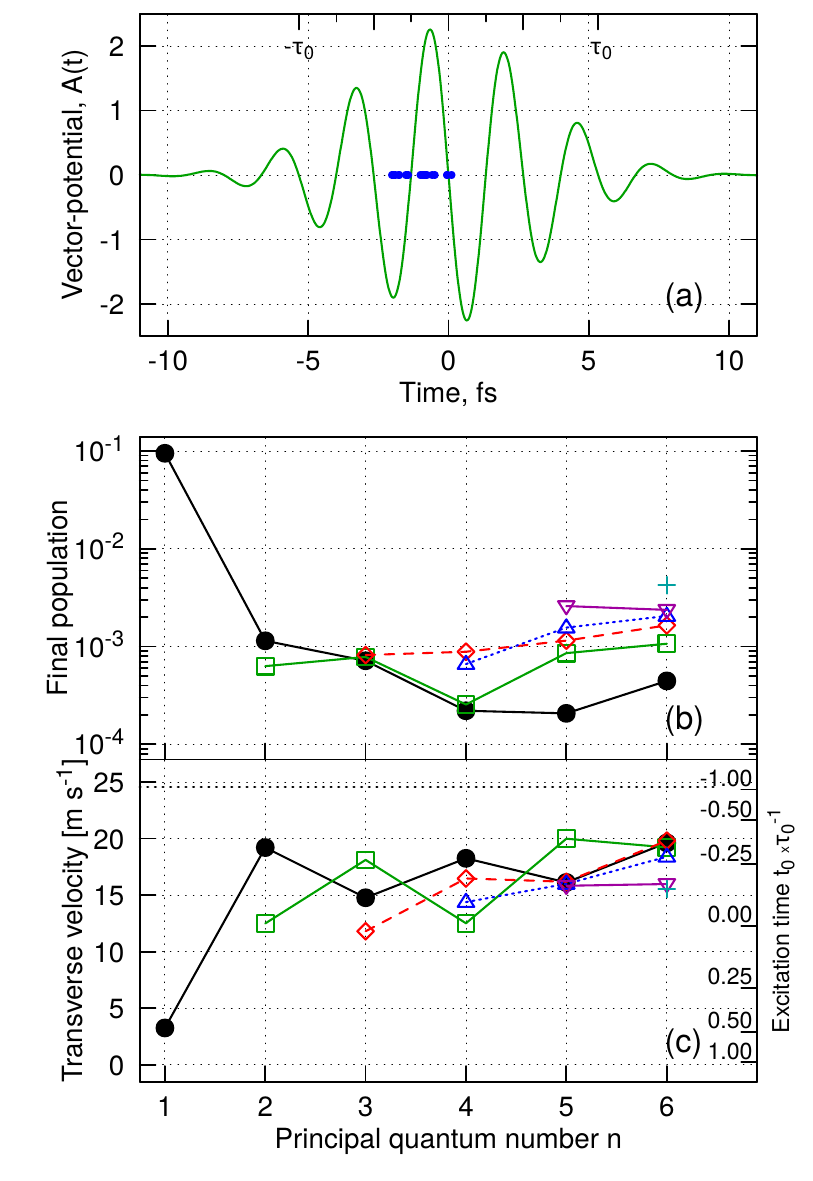} 
  \end{center}
  \caption{
(Color online) Same as Fig.~\ref{fig:w0halfave}, but for the vector-potential
CEP of $\pi/2$.  The results are an equally-weighted average over the initial
positions are within $\pm648$\bohr of the beam half-waist position along $y$.
Also see captions of Figs.~\ref{fig:w0halfave} and \ref{fig:w0half}. See
Fig.~\ref{fig:w0halfavepihalf:times} for the detailed reconstruction of the
excitation times (blue dots on the time axis, panel a).
} \label{fig:w0halfavepihalf}
\end{figure}

\begin{figure}[!tbhp]
  \begin{center}
    \includegraphics[trim=0 0.7cm 0 0,width=\figurewidth]{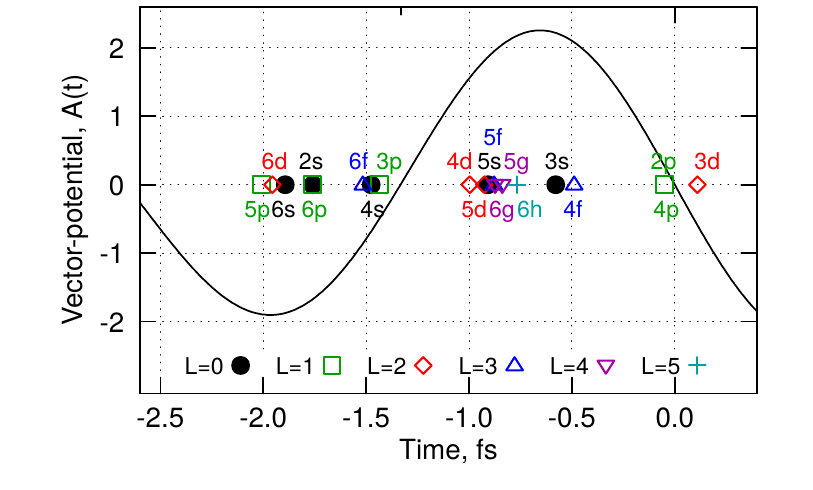} 
  \end{center}
  \caption{
(Color online) Reconstructed excitation times for the data in Fig.~\ref{fig:w0halfavepihalf}.
Also see Fig.~\ref{fig:w0halfave:times} caption.
} \label{fig:w0halfavepihalf:times}
\end{figure}

Up to now, we have considered a short, 2-cycle pulse with cosine CEP for
the vector-potential. It is important to establish that the reconstruction is
applicable to other pulse phases, longer pulses, and for the more realistic,
non-paraxial tightly-focused beam models.

The results for the pulse with sine vector-potential CEP phase (all other
parameters remain the same as is Fig.~\ref{fig:w0halfave}) are shown in
Fig.~\ref{fig:w0halfavepihalf}.  This pulse leaves $\approx\!9.5\%$ of the atoms
in the $1s$ ground state, $\approx\!2.5\%$ in Rydberg states with $n\!\le\!6$, and at
least $2.5\%$ in higher Rydberg states. The final transverse velocity of the
$1s$ ground state is $\approx\!3.3$~m~s$^{-1}$. The reconstructed excitation times
are shown in Fig.~\ref{fig:w0halfavepihalf:times}. These results are
qualitatively similar to Figs.~\ref{fig:w0halfave},\ref{fig:w0halfave:times} of
the main text: The excitation is confined to the single laser cycle, immediately 
preceeding the peak of the laser field. The low-$n$ states tend to be formed
later within the cycle, although the apparent preference is weaker than for the
zero CEP (Fig.~\ref{fig:w0halfave:times}).

\begin{figure}[!tbhp]
  \begin{center}
    \includegraphics[trim=0 0.7cm 0 0,width=\figurewidth]{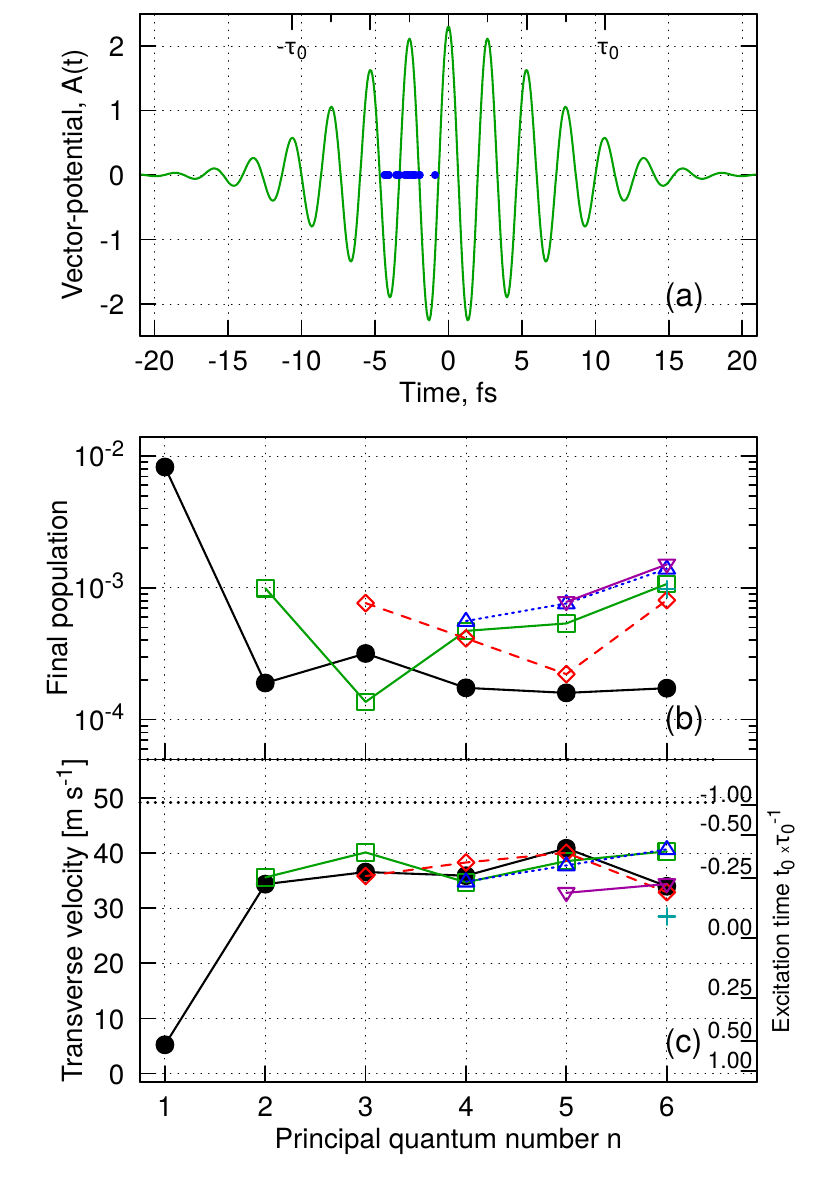} 
  \end{center}
  \caption{
(Color online) Same as Fig.~\ref{fig:w0halfave}, but for the pulse duration of
$\tau_0=440$\au ($\approx\!10.64$~fs).  The results are an equally-weighted
average over the initial positions are within $\pm648$\bohr of the beam
half-waist position along $y$.  Also see caption of Figs.~\ref{fig:w0halfave}
and \ref{fig:w0half}, and Fig.~\ref{fig:w0halfavelong:times} for the
excitation-time reconstruction.
} \label{fig:w0halfavelong}
\end{figure}

\begin{figure}[!tbhp]
  \begin{center}
    \includegraphics[trim=0 0.7cm 0 0,width=\figurewidth]{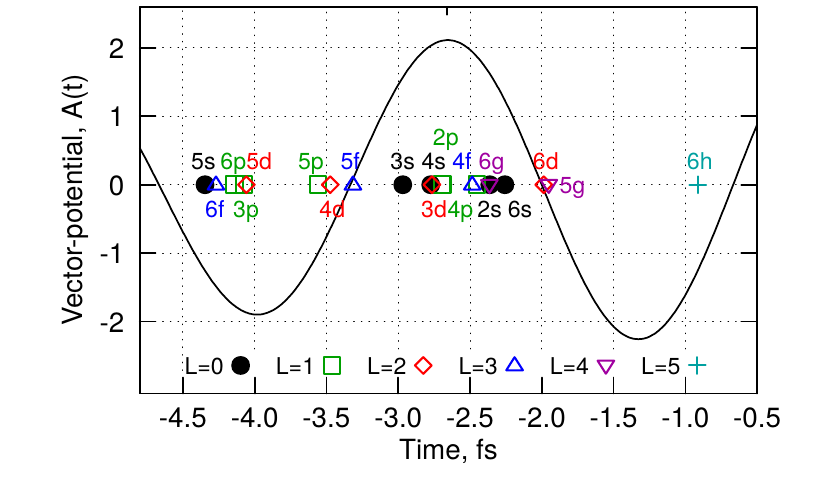} 
  \end{center}
  \caption{
(Color online) Reconstructed excitation times for the data in Fig.~\ref{fig:w0halfavelong}.
Also see Fig.~\ref{fig:w0halfave:times} caption.
} \label{fig:w0halfavelong:times}
\end{figure}

The results for the 440\au ($\approx10.64$~fs) FWHM, cosine CEP pulse (all
other parameters same as in Fig.~\ref{fig:w0halfave}) are shown in
Fig.~\ref{fig:w0halfavelong}. The longer pulse leads to a stronger depletion,
with only $0.8\%$ of the atoms surviving in the $1s$ ground state. The fraction
of the Rydberg states remains essentially unchanged, with $1.2\%$ of the atoms
remaining in $m\!=\!0$ states with $n\!\le\!6$, and at least $1.3\%$ in higher
states.  Due to the longer pulse duration, the ground state now reaches the
higher transverse velocity of $\approx\!5.2$~m~s$^{-1}$, $11\%$ of the free
electron limit ($49.1$~m~s$^{-1}$).  Our analysis still concentrates on the
$m\!=\!0$ states, which dominate the $n\!\le\!6$ population.  The reconstructed
excitation times are shown in Fig.~\ref{fig:w0halfavelong:times}. These values
follow the same pattern as before, with the excitation preferentially occurring
within a single laser cycle. In contrast to the results seen with the shorter,
2-cycle pulses, the excitation occurs earlier, approximately one cycle before
the peak of the envelope is reached. The preference for the late formation
of the low-$n$ states is now essentially absent.

\begin{figure}[!tbhp]
  \begin{center}
    \includegraphics[trim=0 0.7cm 0 0,width=\figurewidth]{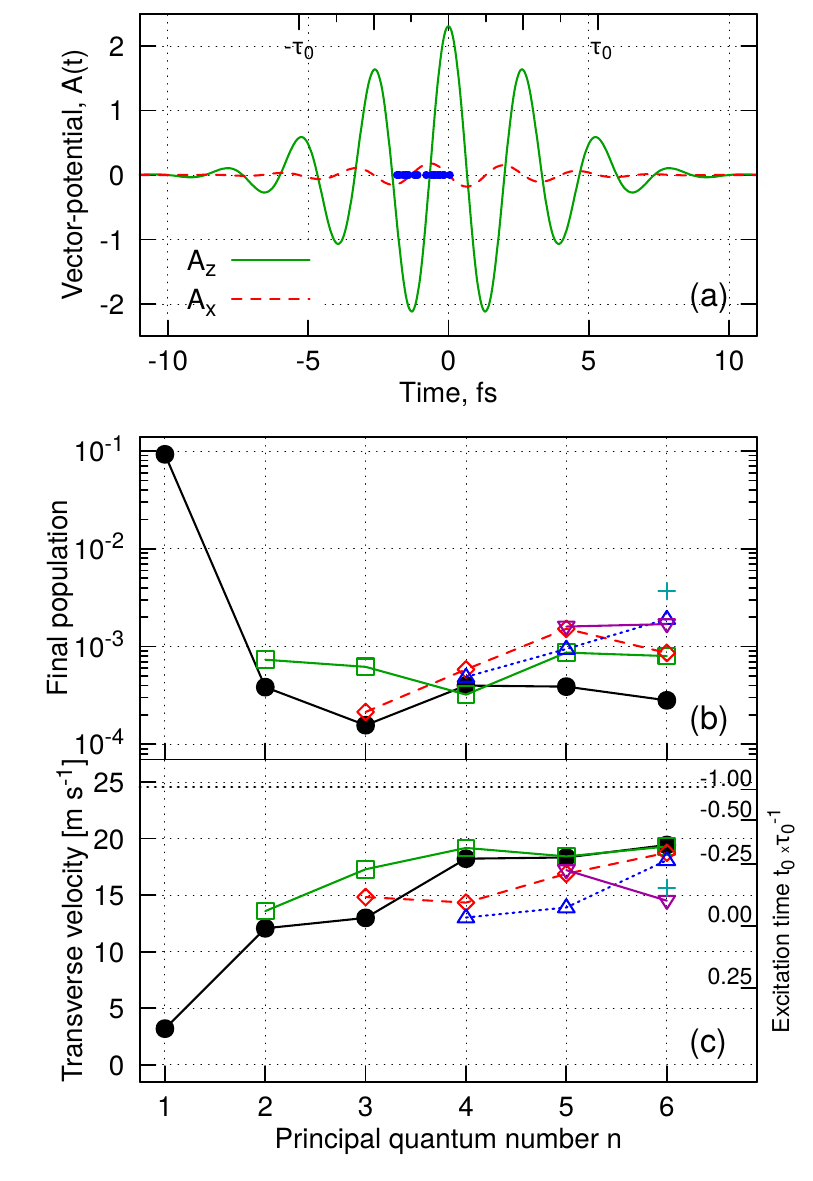} 
  \end{center} 
  \caption{
(Color online) Same as Fig.~\ref{fig:w0halfave}, but for the initial position
displaced along the laser polarization ($z$) direction $(0,0,w_0/2\pm648)$.
Non-paraxial effects are no longer negligible for this beam position, see
Eq.~\eqref{eqn:Ax}. (a) Vector-potential in the along the polarization
(green solid line) and propagation (red dashed line) directions.  Also see
caption of Figs.~\ref{fig:w0halfave} and \ref{fig:w0half}, and 
Fig.~\ref{fig:w0halfavepol:times} for reconstructed excitation times.
} \label{fig:w0halfavepol}
\end{figure}

\begin{figure}[!tbhp]
  \begin{center}
    \includegraphics[trim=0 0.7cm 0 0,width=\figurewidth]{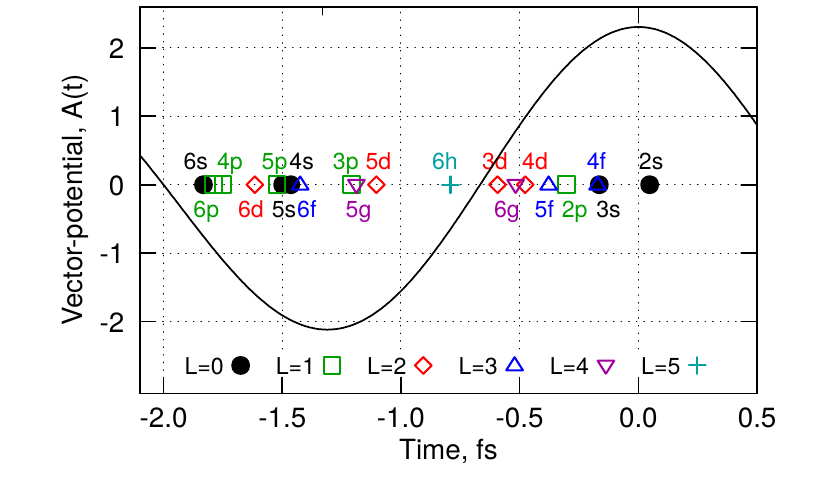} 
  \end{center}
  \caption{
(Color online) Reconstructed excitation times for the data in Fig.~\ref{fig:w0halfavepol}.
Also see Fig.~\ref{fig:w0halfave:times} caption.
} \label{fig:w0halfavepol:times}
\end{figure}

Our final test case uses the same pulse and beam parameters as
Fig.~\ref{fig:w0halfave}, but considers an initial position displaced along the
field polarization direction $z$. For this geometry, the non-paraxial
contributions to the vector-potential become significant (see Eq.~\ref{eqn:Ax}
and red dashed line in Fig.~\ref{fig:w0halfavepol}a).  Again, the numerical
results are averaged along a $\pm648$\bohr line in the maximum
intensity-gradient direction ($z$).  The ground-state survival probability is
unchanged at $9\%$, compared to Fig.~\ref{fig:w0halfave}.  The excitation
probabilities remain similar, at $\approx\!1.8\%$ for $n\!\le\!6$ and at least
$1.7\%$ for higher Rydberg states with $m\!=\!0$. At the same time, the laser
field now has a non-paraxial component in the propagation direction, which
enables efficient excitation of $m\!\ne\!0$ states.  The transverse velocity of
the $1s$ ground state remains unchanged at $\approx\!  3.2$~m~s$^{-1}$.
Finally, the reconstructed excitation times, shown in
Fig.~\ref{fig:w0halfavepol:times}, follow the same pattern as before. It 
therefore appears that the non-paraxial effects do not play a significant
role for our field parameters.

\begin{figure}[!tbhp]
  \begin{center}
    \includegraphics[trim=0 0.7cm 0 0,width=\figurewidth]{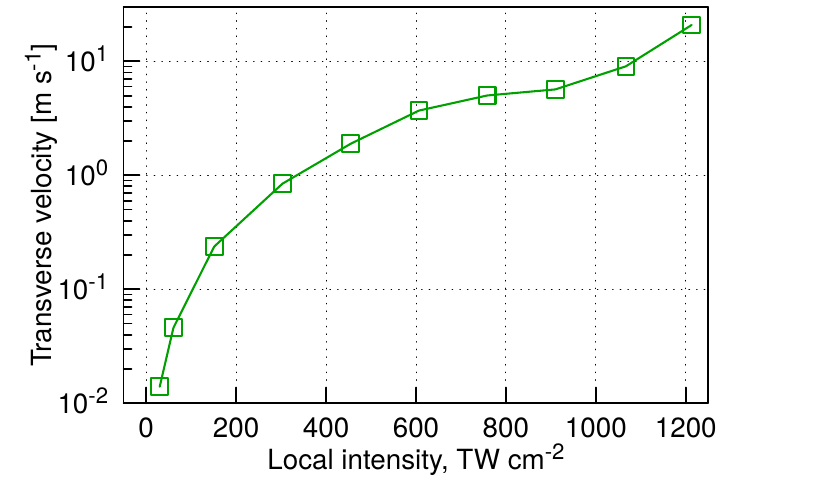} 
  \end{center}
  \caption{
(Color online) Final outward transverse velocity for the hydrogen $1s$ ground state
as a function of the peak intensity at the center of the beam. Pulse parameters and
the initial conditions are given in the Fig.~\ref{fig:ground-polar} caption.
} \label{fig:velocity-1s}
\end{figure}

\clearpage

\bibliography{bib_ei}

\clearpage

\makeatletter\@input{xx-main.tex}\makeatother